\documentclass{aa}  
\usepackage{graphicx}
\newcommand{\teff}{$T_{\rm eff}$}
\newcommand{\kms}{km~s$^{-1}$}
\newcommand{\ebv}{E$(B-V)$}
\newcommand{\gsim}{\;\lower.6ex\hbox{$\sim$}\kern-7.75pt\raise.65ex\hbox{$>$}\;}
\newcommand{\lsim}{\;\lower.6ex\hbox{$\sim$}\kern-7.75pt\raise.65ex\hbox{$<$}\;}

\begin{document}
   \title{The chemical abundance of the
   very metal rich old Open Clusters NGC 6253 and NGC 6791\thanks{
   Based on observations made at ESO	
   telescopes under programmes 67.D-0018, 169.D-0473 and with the Italian Telescopio Nazionale Galileo (TNG)
operated on the island of La Palma by the Fundaci\' on Galileo Galilei of the
INAF (Istituto Nazionale di Astrofisica) at the Spanish Observatorio del Roque
de los Muchachos of the Instituto de Astrofisica de Canarias}}

   \author{
   E. Carretta\inst{1},
   A. Bragaglia\inst{1} \and
   R.G. Gratton\inst{2}
          }

   \offprints{E. Carretta}

   \institute{INAF-Osservatorio Astronomico di Bologna,
              via Ranzani 1, I-40127 Bologna, Italy\\
              \email{eugenio.carretta, angela.bragaglia @oabo.inaf.it}
         \and
          INAF-Osservatorio Astronomico di Padova,
              vicolo Osservatorio 5, I-35122 Padova, Italy\\
              \email{raffaele.gratton@oapd.inaf.it}     }

   \date{}

\abstract
{}  
{In the framework of a project aiming at deriving in a homogeneous
way the properties (age, distance, reddening and detailed chemical
abundances) of a large sample of old open clusters, we present here
the metal abundance and the abundance ratios of light (C, N, O,
Na, Mg, Al, Si, Ca, Ti) and heavier (Cr, Mn, Ni, Ba, Eu) elements in
the galactic open clusters NGC~6253 and  NGC~6791.}
{We performed spectrum synthesis of selected lines on high resolution spectra of
four red clump stars in NGC~6253, taken with the UVES and FEROS spectrographs. 
We also
determined abundances of the same elements for four red clump stars in NGC~6791,
observed with SARG, for which we had derived the atmospheric parameters and the
iron, carbon and oxygen abundances in a previous paper  (Gratton et al. 2006).}
{The average metallicity of NGC~6253 is [Fe/H]$=+0.46$ (rms = 0.03 dex, 
systematic error = 0.08 dex),
obtained by extensive spectral synthesis of  Fe lines. This intermediate age
cluster closely resembles the old open cluster NGC~6791, as far as the 
chemical composition is concerned. 
C, N, O  do not show any significant abundance scatter; they are
underabundant with respect to the solar values both in NGC~6253 and
NGC~6791. We also 
find no evident
star-to-star scatter in any of the elements measured in both clusters, with the
possible exception of Na in NGC~6791.
The two clusters show very similar abundances, except for Mg,
overabundant in NGC~6791 and not in NGC~6253. Both have solar scaled
$\alpha$-elements abundances. We have compared our abundance
ratios with literature values for disk giants and dwarfs and bulge giants,
finding a general good agreement with the run of elemental ratios with [Fe/H]
of disk objects.}
{}

\keywords{ Stars: Evolution - Stars: Abundances - Galaxy: Disk - 
Open clusters and associations: General - Open clusters and associations:
individual (NGC~6253, NGC~6791) }

\titlerunning{NGC~6253 and NGC~6791}
\authorrunning{Carretta et al.}

\maketitle

%

\begin{table*}
\centering
\caption{
Information on the target stars. ID, $V$, $B-V$ are taken from Bragaglia et al.
(1997) and $J$, $K$ from 2MASS; S/N is measured near 6100 \AA, the radial
velocity (RV) is heliocentric, "bin?" in Notes means non member or
binary.}
\begin{tabular}{cccccccccccl}
\hline
 ID  &	 RA 	   & Dec	& Date & exptime  & S/N & $V$     & $B-V$  &$J$	   &$K$	    &RV     &Notes\\	  
     & ($h$ ~$m$ ~$s$)  &($\degr$ ~~$\arcmin$ ~~$\arcsec$)  &      & (s)      &     &	   &	  &  &  &(\kms) &\\
\hline
2509 & 16 59 15.91 &-52 42 26.7 & 2002-07-16 & 1800 	     &120 &12.685 &1.314 &10.206 &9.428 &-28.71& UVES\\
2885 & 16 58 53.28 &-52 41 54.4 & 2002-07-19 & 2400 	     &180 &12.656 &1.352 &10.121 &9.296 &-28.13& UVES\\
4510 & 16 59 06.03 &-52 39 56.6 & 2002-07-19 & 1800 	     &150 &12.759 &1.296 &10.279 &9.470 &-27.44& UVES\\
2508 & 16 59 03.78 &-52 42 25.9 & 2001-04-26 & 2$\times$4200 & 85 &12.685 &1.284 &10.252 &9.450 &-20.60& FEROS, bin?\\
3595 & 16 59 05.80 &-52 41 04.9 & 2001-04-25 & 2$\times$4200 & 85 &12.388 &1.292 &~9.951 &9.192 &-28.76& FEROS\\
\hline
\end{tabular}
\label{tab-data}
\end{table*}

\section{Introduction}

This is the fifth paper of the spectroscopic part of the  Bologna Open Cluster
Chemical  Evolution (BOCCE) project (see Bragaglia \& Tosi 2006, Carretta et al.
2004, 2005), aimed at deriving  with the best attainable homogeneity  
reddening, age and metallicity for a sample of old OCs in order to define  the
metallicity distribution in the disk and its (possible) evolution with time.  We
concentrate here on the two very metal rich clusters NGC~6253 and NGC~6791,  and
present a first detailed abundance analysis based on high S/N, high resolution 
spectra of red clump stars in NGC~6253 and  the follow-up of the work in 
Gratton et al. (2006, hereafter G06) on NGC~6791.

We selected the cluster NGC~6253 as relevant in determining the presence and
slope of the disk radial abundance gradient since it is one of the known OC with
smaller galactocentric distance and its metallicity is very high. We have
recently re-analyzed (Bragaglia \& Tosi 2006) the cluster photometric data  with
updated evolutionary tracks and confirmed the findings in  Bragaglia et al.
(1997). From comparison of synthetic CMDs to the observed ones, NGC~6253 has
R$_{GC}$ = 6.6 kpc, Z=0.05 and age=3 Gyr. Other papers on this cluster were
published by Piatti et al. (1998) and  Sagar et al. (2001) based on broad-band
photometry, by Twarog et al. (2003 - with results recently revised in
Anthony-Twarog et al. 2007) based on Str\"omgren photometry, by Carretta et al.
(2000) and Sestito et al. (2007)  based on high resolution spectroscopy.
The high metallicity of NGC~6791 has been acknowledged for some time (e.g.,
Peterson \& Green 1998 and Chaboyer et al. 1999, based on spectroscopy and
photometry, respectively) and has been confirmed by several very recent works:
our own paper (G06, high resolution optical spectroscopy),  Carraro et al.
(2006, high resolution optical spectroscopy), Origlia et al. (2006, high
resolution infrared spectroscopy) and Anthony-Twarog et al. (2007, Str\"omgren
photometry). More details on these works will be presented in Section 5.

Cool stars, like the ones studied here, and high metal abundance conspire to
produce very crowded spectra  that are better examined by spectrum synthesis
than with the classical abundance analysis based on equivalent widths, even at
the quite high resolution of the FEROS and UVES spectra, R$\simeq$45000,
considered in the present paper.  The present analysis is hence done in the same
way as the one for NGC~6791 (G06), i.e. employing extensive synthesis of
selected features.

The paper is organized as follows: the data are described in Sect. 2,
metallicity determination is discussed in Sect. 3, while abundances of other
elements are presented in Sect. 4 for NGC~6253 and 
for NGC~6791. Sect. 5 presents a comparison to literature data on these
two OCs and field stars; a discussion and summary are given in
Sect. 6. 

\section{Observations and data reduction}

\begin{figure}
\centering
\includegraphics[bb=100 200 500 690, clip,scale=0.6]{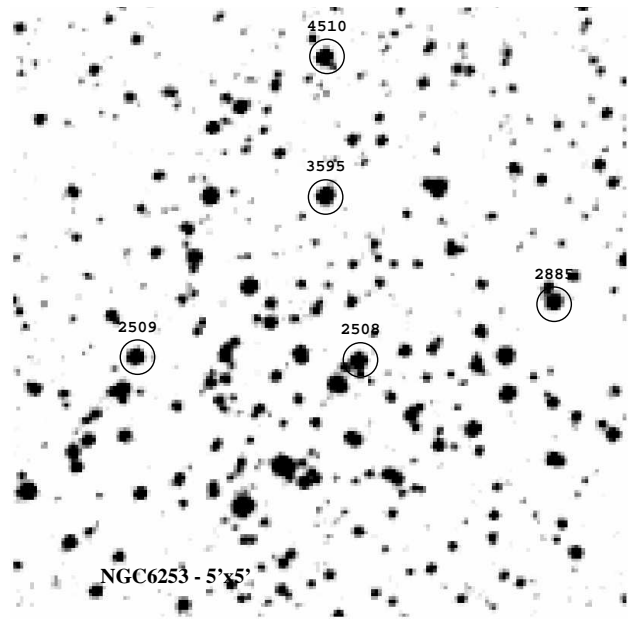}
\caption{Map of NGC~6253 (5 $\times$ 5 arcmin$^2$, North up, East left) with
the observed stars.} 
\label{fig-map}
\end{figure}

\begin{figure}
\centering
\includegraphics[bb=60 190 380 500, clip,scale=0.75]{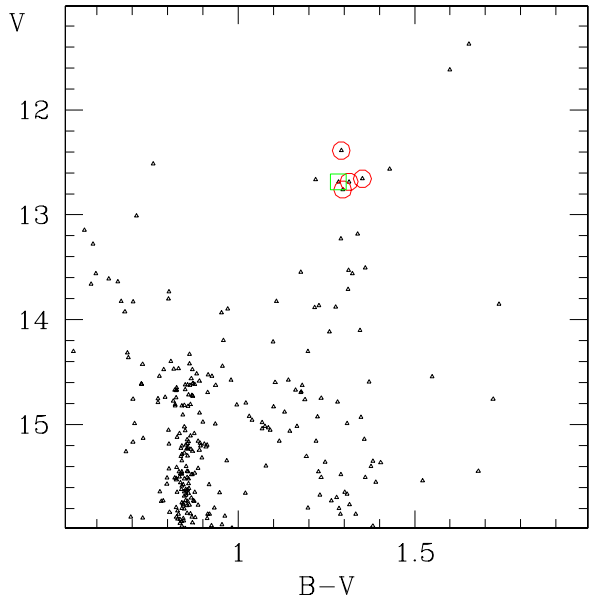}
\caption{$V,B-V$ CMD of the central region of
NGC~6253 (from Bragaglia et al. 1997), with the observed stars
shown as larger symbols; circles indicate single, member stars
and the square the probable binary (or non member) star.} 
\label{fig-cmd}
\end{figure}

\paragraph{NGC~6253:} 
As done for the other clusters of the BOCCE sample, our targets  in NGC~6253
were chosen among Red Clump (RC) stars, using the photometry by Bragaglia et al.
(1997). The targets lie near the cluster centre and their positions are shown in
Fig. \ref{fig-map}.  Fig. \ref{fig-cmd} displays the cluster CMD with the five
targets indicated by larger symbols. Identifications, celestial coordinates
(J2000), photometry and  details on the observations are presented in
Table~\ref{tab-data}. 

\begin{figure}
\centering
\includegraphics[bb=40 160 570 700, clip,scale=0.47]{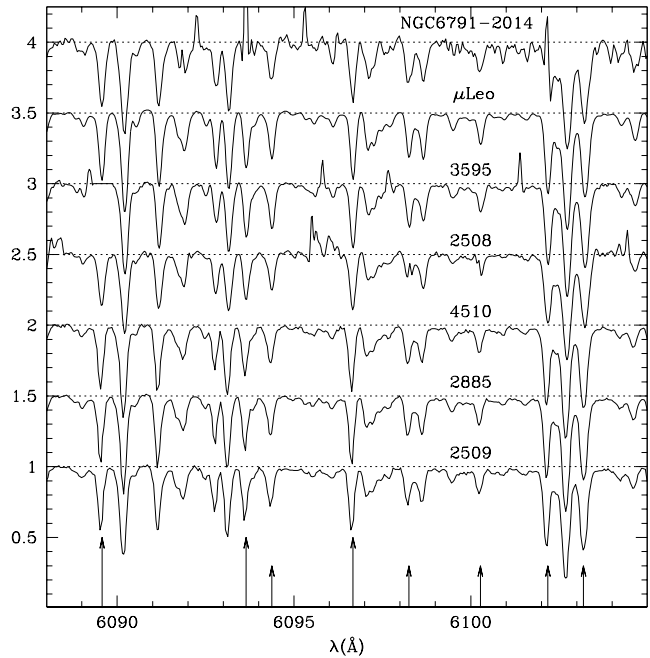}
\caption{Small fraction of the spectra of the 5 target stars in NGC~6253, of
the field giant star $\mu$ Leo, and of a RC star in NGC~6791; the spectra
are offset for clarity. The arrows point to
Fe {\sc i} lines; the longer ones indicate lines used in our synthesis. The Fe
line near 6090 \AA \ for star 3595 has been excised because of a defect.} 
\label{fig-spe}
\end{figure}

Observations for stars 2508 and 3595 were obtained with FEROS, a high resolution
fiber-fed spectrograph at the time mounted on the 1.5m ESO telescope in La
Silla, on 2001 April 25 and 26 (programme 67.D-0018). The wavelength range is
3700-8600 \AA, at R=48000. Two exposures were obtained for each star and the
individual spectra were reduced using the data reduction package installed at
the telescope  and later summed.  Stars 2509, 2885, and 4510 were observed 
using UVES on the ESO VLT (Unit 2),  on 2002 July 16 and 19 (as a backup during
programme 169.D-0473); a single exposure was obtained for each star. The spectra
cover the wavelength ranges 3560-4840 and 5710-9320 \AA, at R=43000. They were
reduced with the standard pipeline. 

We measured radial velocities (RVs) for each individual spectrum using several
tens of metallic lines; the resulting heliocentric RVs are listed  
in Table~\ref{tab-data}; their errors are less than about 1 km~s$^{-1}$.

\paragraph{NGC~6791:} 
Four RC stars in the very old (about 9 Gyr, King et al. 2005), very metal rich
([Fe/H]$=+0.47$, according to G06) open cluster NGC~6791 were obtained with SARG
at the 3.5m Italian telescope Galileo at a slightly lower resolution
(R$\simeq$30000) and S/N (60 on average) than those considered for 
NGC~6253. The grating used covers the wavelength  range 4595-6170 and 6240-7900
\AA; however only the central interval ($\sim$5000-6500 \AA) was considered
of good enough quality to be used
in the analysis. We recall here that the RC in NGC~6791 is -apparently-
much fainter than that of NGC~6253, due to the larger distance; typical
magnitudes and colours are
$V\simeq14.6$,  $B-V \simeq 1.4$ and $V-K \simeq 3.1$.  Details on observations
and reductions, coordinates and magnitudes, etc. can be found in G06. 

\section{Abundance analysis}

\subsection{Stellar parameters}

The required atmospheric parameters for the stars in  NGC~6253 were derived from
the photometry.  Effective temperatures (\teff) and surface gravities ($\log
g$)  were derived from the $V-K$ colour calibrations for temperature and
bolometric corrections by Alonso et al. (1999), using the
$B$, $V$ visual photometry by Bragaglia et al. (1997) and the 
$J$, $K$ infrared photometry by 2MASS (Skrutskie et al. 2006).   
In
the derivation of atmospheric parameters  we adopted  $(m-M)_0=11.0$, \ebv=0.23,
mass=1.4 M$_\odot$ from Bragaglia \& Tosi (2006). We also assumed [Fe/H]=+0.2
since the latter represents the high-metallicity limit for the Alonso et al.
calibration. However, the negligible dependency of the $V-K$ colour on
metallicity ensures that we are not committing a significant error in the case
of the stars of NGC~6253, as also stated by Alonso et al. (sect.
2.3), who consider $V-K$ as the best temperature indicator for giant stars,
since an error of 0.5 dex in metallicity translates in and error of at most
0.7\% in $T_{\rm eff}$.
The microturbulent velocities ($v_t$) were computed
using the relation  with $\log g$ presented in Carretta et al. (2004).  The
resulting atmospheric parameters are listed in Table~\ref{t:paratm}.

We estimated the error in \teff \ and $\log g$ due to our choice of distance and
reddening as follows. The ranges covered in literature for these
parameters are rather small: $(m-M)_0$=11.0$\pm$0.25 and \ebv=0.23$\pm$0.03. A
variation of 0.03 in \ebv \ translates into about 70~K when \teff \ is derived
by $V-K$, while a variation of 0.25 in distance modulus translates into 0.1 dex
in $\log g$. These are the major error sources in both these parameters; we
do not have any indication of errors larger than those estimated from our
abundance analysis (see Sect. 3.2).

Atmospheric parameters for the four RC star in NGC~6791 are taken from G06 (their
Table 3) where   a detailed description can be found; we report them in
Table~\ref{t:paratm} for convenience. They were derived in the same way as in
NGC~6253, adopting $(m-M)_V$=13.45 and \ebv=0.15 (as average of the literature
values) and mass 0.9 M$_\odot$ (for an age of 9 Gyr, e.g., King et al. 2005).

\begin{table}
\centering
\caption{Atmospheric parameters for the stars
observed in NGC~6253 and NGC~6791.}
\begin{tabular}{cccccrrc}
\hline
 ID  & $T_{\rm eff}$ &$\log g$ &$v_t$   \\	    
     & K             &         &\kms    \\
\hline
\multicolumn{4}{c}{NGC 6253}\\
2509 &4518 &2.45 &1.18 \\
2885 &4438 &2.39 &1.19 \\
4510 &4492 &2.47 &1.18 \\
2508 &4536 &2.47 &1.18 \\
3595 &4568 &2.37 &1.19 \\
\multicolumn{4}{c}{NGC 6791}\\
2014 &4463 &2.30 &1.05 \\
3009 &4473 &2.33 &1.05 \\
3019 &4468 &2.35 &1.05 \\
SE49 &4512 &2.32 &1.05 \\
\hline
\end{tabular}
\label{t:paratm}
\end{table}

\begin{table*}
\centering
\caption{Iron abundances in RC stars of NGC~6253. 
For each Fe line synthesized in the 4 member stars we show the
wavelength, the adopted EP and $\log gf$, the individual values of  $\log n$ 
(cols. 4-7); the cluster average and rms for each line are in cols. 8-9. 
}
\begin{tabular}{ccccccccc}\\
\hline
 Wavelength &EP$_{low}$&$\log gf$&$\log n$ &$\log n$ &$\log n$ &$\log n$ &$\log n$ & \\
 \AA        &eV&         &2509     &2885     &4510     &3595	 & mean    & rms \\
\hline
\multicolumn{9}{c}{Fe {\sc i}}\\
 5560.22 &4.43 &-1.10 &      &      &	   & 7.94 &  7.94 &     \\
 5577.03 &5.03 &-1.49 &      &	    &	   & 7.89 &  7.89 &	\\
 5618.64 &4.21 &-1.34 &      &	    &	   & 7.94 &  7.94 &	\\
 5619.61 &4.39 &-1.49 &      &	    &	   & 7.94 &  7.94 &	\\
 5651.48 &4.47 &-1.79 &      &	    &	   & 7.94 &  7.94 &	\\
 5661.35 &4.28 &-1.83 &      &	    &	   & 7.94 &  7.94 &	\\
 6065.49 &2.61 &-1.53 & 8.04 & 7.94 & 7.92 & 7.84 &  7.94 &0.08 \\
 6078.50 &4.80 &-0.48 & 8.14 & 8.14 & 8.09 & 8.16 &  8.13 &0.03 \\
 6079.02 &4.65 &-0.95 & 7.99 & 8.12 & 7.84 & 8.12 &  8.02 &0.13 \\
 6089.57 &4.58 &-1.28 & 7.84 & 7.79 & 7.76 &	  &  7.80 &0.04 \\
 6093.65 &4.61 &-1.32 & 7.94 & 7.89 & 7.76 & 8.04 &  7.91 &0.12 \\
 6094.38 &4.65 &-1.56 & 7.93 & 7.84 & 7.82 &	  &  7.86 &0.06 \\
 6096.67 &3.98 &-1.77 & 7.94 & 8.14 & 8.04 & 8.02 &  8.04 &0.08 \\
 6097.09 &2.18 &-5.01 & 7.84 & 7.84 & 7.84 &	  &  7.84 &0.00 \\
 6151.62 &2.18 &-3.30 & 7.94 & 8.14 & 7.94 & 8.14 &  8.04 &0.12 \\
 6165.36 &4.14 &-1.50 & 7.94 & 8.12 & 8.02 & 7.94 &  8.01 &0.09 \\
 6232.65 &3.65 &-1.22 & 8.04 & 8.14 & 7.94 & 7.94 &  8.02 &0.10 \\
 6246.33 &3.60 &-0.73 & 7.99 & 7.94 & 7.92 & 8.02 &  7.97 &0.05 \\
 6252.56 &2.40 &-1.69 & 7.89 & 7.94 & 7.84 & 8.12 &  7.95 &0.12 \\
 6270.23 &2.86 &-2.46 & 8.12 & 8.14 & 7.94 &      &  8.07 &0.11 \\
 6297.80 &2.22 &-2.74 & 8.04 & 8.12 & 8.04 & 8.12 &  8.08 &0.05 \\
 6301.51 &3.65 &-0.72 & 8.04 & 8.04 & 8.04 & 7.96 &  8.02 &0.04 \\
 6392.54 &2.28 &-3.97 & 8.04 & 8.12 & 7.93 & 7.94 &  8.01 &0.09 \\
 6393.61 &2.43 &-1.43 &      & 7.74 &      &      &  7.74 &	\\
 6411.66 &3.65 &-0.60 & 8.14 & 8.14 & 8.14 & 8.14 &  8.14 &0.00 \\
 6421.36 &2.28 &-2.03 & 7.79 & 7.84 & 7.76 & 7.97 &  7.84 &0.09 \\
 6481.88 &2.28 &-2.98 & 8.24 &      & 8.14 &      &  8.19 &0.07 \\
 6498.95 &0.96 &-4.66 &      &	    &	   & 8.14 &  8.14 &	\\
 6574.25 &0.99 &-5.00 &      &      &      &	  &  	  &	\\
 6593.88 &2.43 &-2.42 & 8.09 & 8.14 & 8.14 & 8.04 &  8.10 &0.05 \\
 6608.04 &2.28 &-3.96 & 7.96 & 8.12 & 8.02 & 8.04 &  8.04 &0.07 \\
 6609.12 &2.56 &-2.69 & 8.14 &      & 8.13 & 8.14 &  8.14 &0.01 \\
 6703.58 &2.76 &-3.00 &      & 7.79 &      &      &  7.79 &	\\
 6713.75 &4.80 &-1.41 & 7.89 & 7.96 & 7.96 &	  &  7.94 &0.04 \\
 6725.36 &4.10 &-2.21 & 7.94 & 8.14 & 8.04 & 8.02 &  8.04 &0.08 \\
 6726.67 &4.61 &-1.10 & 7.94 & 8.04 & 7.99 & 8.12 &  8.02 &0.08 \\
\hline
\multicolumn{9}{c}{Fe {\sc ii}}\\
 6149.25  &3.89 &-2.73 & 7.49 & 7.79 & 7.69 & 8.09 &  7.77 &0.25 \\
 6247.56  &3.87 &-2.33 & 7.89 & 7.99 & 7.89 & 8.09 &  7.97 &0.10 \\
 6369.46  &2.89 &-4.21 & 7.91 & 7.99 & 7.89 & 8.07 &  7.97 &0.08 \\
 6416.93  &2.89 &-2.70 & 7.89 & 7.89 & 7.87 & 7.89 &  7.89 &0.01 \\
 6432.68  &2.89 &-3.58 & 8.07 & 8.24 & 8.09 & 8.09 &  8.12 &0.08 \\
 6456.39  &3.90 &-2.10 & 8.19 & 8.29 & 8.19 & 8.11 &  8.20 &0.07 \\
\hline
\end{tabular}
\label{tab-fe}
\end{table*}

\begin{figure*}
\centering
\includegraphics[bb=30 180 580 500, clip,scale=0.7]{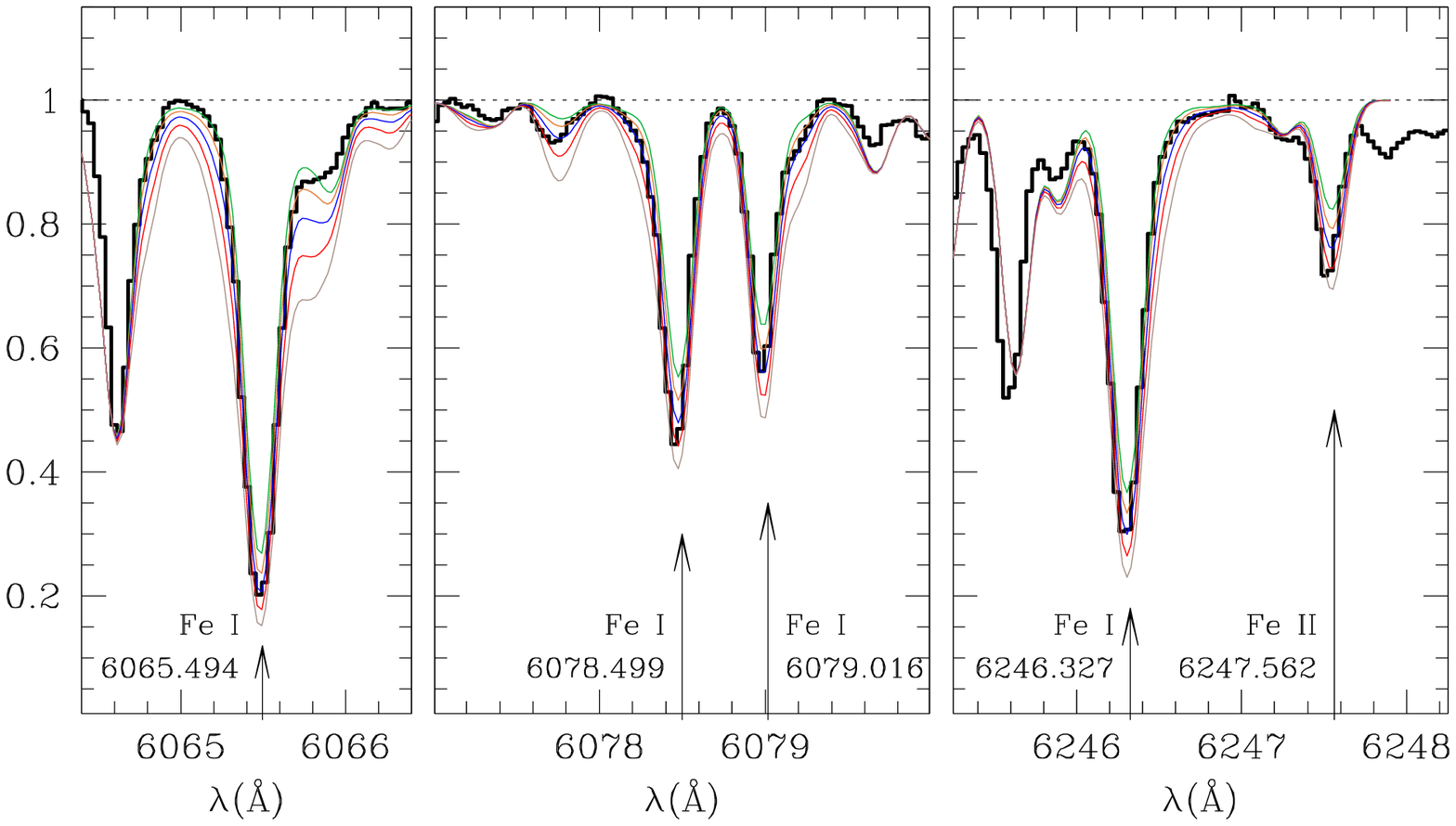}
\caption{Examples of the synthesis of four Fe {\sc i} and one Fe {\sc ii} lines
for star 2885, where the thick line is the observed spectrum.
The synthetic spectra (thin lines) were computed with the following
parameters $T_{\rm eff}$=4438~K, $\log g$=2.39, $v_t$=1.19 \kms, and
[Fe/H]= 0.0, +0.2, +0.4, +0.6, +0.8 from top to bottom, respectively.
} 
\label{fig-sint}
\end{figure*}

\subsection{Errors}
Three kind of errors have to be considered:

{\em Internal errors} (the truly random component), that will show up in
star-to-star comparisons;

{\em Global errors} (originating e.g., from the adopted atmospheric parameters), 
that affect all stars in a cluster in the same way;

{\em Scale errors} (originating e.g., from the solar reference
composition, the un-adequacy of model atmospheres, the impact of non-LTE
effects), that affect all analyses made in the BOCCE program.

These three kind of errors should be considered separately, because they are
estimated using different techniques, and because they have a different impact
on the discussion.

Internal errors are related to the star-to-star scatter. Up to now we have found
very little evidence of any intrinsic scatter in our analysis of
open cluster stars (at variance with the situation of globular cluster stars). 
Hence, a good check on the accuracy of the internal errors is then the observed
star-to-star scatter, or for an individual star the line-to-line scatter.

Global (cluster) errors  are the major matter of concern in the
BOCCE program, because our goal is to use the open clusters as tracers of
Galactic chemical evolution. While residual errors after averaging results for
individual stars are not entirely negligible, the major source of these global
errors are the adopted atmospheric parameters and their dependence
on the interstellar reddening and the distance modulus,  derived from
photometry, mainly (but not exclusively) from our group. The main problem is the
possible degeneracy existing between reddening and metallicity in photometric
analyses. However,   the cluster metallicity has a dependence on the assumed
reddening that is exactly the opposite in the case of a CMD and a spectrum. In
the first case,  the highest the reddening, the most metal poor the cluster
should be; in the second case the adoption of a higher reddening means assuming
that the stars are warmer, which leads to a higher metal abundance from the
analysis of the spectra. Hence, it is very important that results from the
analysis of the CMDs are consistent with those obtained from the spectroscopic
analysis.

Finally, we will leave out scale errors for now, not because
we believe they are not present, but because they affect only at
second order the discussion of the results in the whole BOCCE program,
insofar our assumptions (e.g., line lists, model atmospheres, etc) are kept
constant  throughout our series of papers. This is the main reason we define our
procedure as "homogeneous". Furthermore, even if these stars are not the
brightest ones,  we decided to analyze clump stars because they are not very
cool and have a limited range of atmospheric parameters in all old open
clusters. We think that this is much safer than observing the brightest cluster
stars,  that are generally very cool, so that concerns on the adequacy of
model atmospheres or the impact of non-LTE effects are much more severe.

\subsection{Metallicity of NGC~6253 from synthesis of Fe lines}

As in  G06, we  used spectrum synthesis on selected iron lines (Carretta et al.
2004) to derive the metallicity; we considered 36 Fe {\sc i} and 6 Fe {\sc ii}
lines for the analysis (see Table~\ref{tab-fe}). Holding fixed the $T_{\rm
eff}$,  $\log g$ and $v_t$ values appropriate for each star, we extracted the
corresponding models from the grid of model atmospheres by Kurucz (1993) with
the overshooting option switched on; the model we considered have a solar-scaled
mixture of heavy elements. We then generated synthetic spectra in a 2~\AA \
region around each Fe line, by varying  the iron abundances  from [Fe/H]=$-0.2$
to $+1.0$  in steps of 0.2 dex.  The [Fe/H] value  most appropriate for
each individual line was derived by  best fitting (by eye) the synthetic
spectrum to the observed one. This procedure allowed us to supersede the large
uncertainties due to blending and alleviated the problem of correct continuum
positioning in spectra of rather cool, very metal-rich giant stars. Examples of
the fitting are given in Fig. \ref{fig-sint}. 

A detailed description of the line lists used to compute the synthetic spectra
is given in both Carretta et al. (2004) and G06 and it will not be repeated
here. We only remind that line parameters for iron lines (oscillator strengths
and damping broadening in particular) are as in Gratton et al. (2003); they are
the same used in previous studies on abundances of old open clusters
consistently derived by our group.

In G06 we adopted an estimate of 0.16 dex for the  {\it fitting
error}; this includes random errors due to photometric  uncertainties on the
line profile and to the estimate of the local continuum level.
Using this approach, spectrum normalization is obtained by matching "high"
points in the observed spectrum to corresponding "high" points in the synthetic
spectrum. These high points do not necessarily have the intensity expected
for the local continuum since we are using a very extensive line list tested
on the very high resolution and S/N spectrum of the extremely line-rich K giant 
HR~3627 (see Carretta et al. 2004), whose lines are even stronger than
in the program stars. Some uncertainty in this normalization procedure is
clearly present; this is what we call "error in the estimate of the local
continuum level".
However, the present spectra are of much higher quality than those of
NGC~6791, and we consider a better estimate for this
error the value of 0.12 dex, the typical line-to-line rms scatter for
the four program stars.

Results derived for individual Fe lines are listed in Table \ref{tab-fe}, 
expressed as abundances by number\footnote{We use the usual spectroscopic 
notation: log~n(A) is the abundance (by  number) of the element A in the usual
scale where log~n(H)=12; [A/H] is  the logarithmic ratio of the abundances of
elements A and H in the star, minus  the same quantity in the Sun.} 
both for Fe {\sc i} and Fe {\sc ii} lines. 
Average metallicities are summarized in Table~\ref{t:meanfe}, together with the
number of lines used and the rms scatter for each star;  in this Table,
abundances are  referred to the solar abundances adopted in Gratton et al.
(2003): $\log n$(Fe) = 7.54 and 7.49 from neutral and singly ionized lines,
respectively. There is a very small scatter in metallicity and 
no trend as a function of effective temperature.
In Table~\ref{t:meanfe} we also give the metallicity of the four stars in NGC~6791
for immediate comparison (from G06, Table 3).
 
We do not report here the abundances for star 2508, whose spectrum  has an RV
differing from the average of the four others by about 10$\sigma$ and an abundance
lower than the cluster average.  This star could be a field interloper,
notwithstanding its position on the RC, or a binary. The latter possibility is 
supported by the different RV we determined in Carretta et al. (2000) for this
star ($-28.5$ versus the present $-20.6$ \kms), albeit on spectra of worse
quality  and with a probable zero point effect of the order of $-$2
to $-$3 \kms. 
Furthermore, while writing the paper we were informed (S. Desidera and M.
Montalto, private communication based on an on-going work on planet search in
NGC~6253) that star 2508 has a 92 \% probability of being a cluster member,
based on its proper motion. Of course, complete confirmation of binarity 
will require the acquisition of a series of spectra in order to check for
RV variations.
The lower abundance could be explained by veiling from a secondary
component;  given the difference in line intensity between star 2508 and the
other RC cluster stars (about 10 \%), the secondary component should roughly
provide a fraction of light of about 0.1, i.e., should be about 2.5 mag fainter,
in the observed wavelength range. It could then be a main sequence star, just
below the turn-off point. Its contribution makes the primary appear slightly
brighter and bluer than single RC stars; taking this into account the primary
still has magnitude and colour appropriate for a RC star. Given the larger
uncertainties related to  the analysis of this star, we have excluded it from
further discussion.

Due to the quite similar procedure we used for both clusters  (set of line
parameters and model atmospheres, derivation of atmospheric parameters,
spectrum synthesis technique and package), we adopt the same error budget
derived in G06 for NGC~6791 (see their Table 4). The small,
not statistically significant, difference [Fe/H]{\sc
i} - [Fe/H]{\sc ii} $= -0.04 \pm 0.05$ dex we found between Fe {\sc i} and Fe
{\sc ii} would be erased by  choosing  temperatures about  50 K higher or
gravities 0.15 dex lower than those adopted. On the other hand, the trend of
abundances with line excitation is $\Delta$[Fe/H]/$\Delta$E.P. = $-0.013 \pm
0.014$ dex/eV, on average. Although not significant, this indicates that
temperatures should be $lowered$ by $55 \pm 59$ K in order to match the
excitation temperature. The conclusion is that these uncertainties  are well
within those estimated to affect the adopted atmospheric parameters.

In summary, we derive for NGC~6253 a metal abundance [Fe/H]$=+0.46 \pm 0.02 \pm
0.08$, where the first error bar refers to the random part (as estimated from
the star-to-star scatter: rms=0.03 dex) and the second term takes into account
the systematics due to the assumption of reddening, distance modulus,
temperature  scale, etc (see Sect. 3.3 in G06).

\begin{table}
\centering
\caption{Average Fe abundances for stars in in NGC~6253 and NGC~6791, based on
spectrum synthesis. }
\begin{tabular}{ccllcll}
\hline
 ID  & nr & [Fe/H] {\sc i}& $\sigma$ & nr & [Fe/H] {\sc ii}& $\sigma$\\ 
\hline
\multicolumn{7}{c}{NGC 6253}\\
2509 & 26 &   +0.45  &  0.11  & 6  &  +0.42  &  0.24   \\
2885 & 26 &   +0.47  &  0.14  & 6  &  +0.54  &  0.20   \\
4510 & 26 &   +0.42  &  0.12  & 6  &  +0.45  &  0.18   \\
3595 & 27 &   +0.48  &  0.09  & 6  &  +0.57  &  0.08   \\
\multicolumn{7}{c}{NGC 6791}\\
2014 & 26 &   +0.40  &  0.12  & 6  &  +0.41  &  0.11   \\
3009 & 29 &   +0.56  &  0.14  & 6  &  +0.38  &  0.24   \\
3019 & 28 &   +0.45  &  0.11  & 5  &  +0.36  &  0.21   \\
SE49 & 28 &   +0.47  &  0.18  & 6  &  +0.60  &  0.33   \\
\hline
\end{tabular}
\label{t:meanfe}
\end{table}

\begin{table}
\centering 
\caption{Atomic lines used in the synthesis of other elements.
}
\begin{tabular}{ccc}
\hline
$\lambda$ (\AA)& EP$_{low}$ & $\log gf$   \\
\hline
\multicolumn{3}{c}{[O {\sc i}]}  \\
 6300.311 & 0.00 & -9.75  \\ 
\multicolumn{3}{c}{Na {\sc i}}    \\
6154.226 & 2.102 & -1.57  \\ 
6160.747 & 2.104 & -1.26   \\								  
\multicolumn{3}{c}{Mg {\sc i}}   \\ 
6318.717 & 5.108 & -1.94   \\ 
6319.237 & 5.108 & -2.16     \\ 
6319.494 & 5.108 & -2.67   \\ 
7691.553 & 5.574 & -0.65   \\
\multicolumn{3}{c}{Al {\sc i}}  \\
6696.032 & 3.140 & -1.32  \\ 
6698.671 & 3.140 & -1.62  \\					     
\multicolumn{3}{c}{Si {\sc i}}   \\
5666.686 & 5.616 & -1.81  \\
5684.493 & 4.950 & -1.65  \\
5690.425 & 4.930 & -1.87  \\
6155.693 & 5.619 & -2.46  \\
6721.844 & 5.860 & -1.21  \\
7932.348 & 5.964 & -0.47  \\				      
\multicolumn{3}{c}{Ca {\sc i}}   \\  
6156.023 & 2.521 & -2.45  \\ 
6161.297 & 2.523 & -1.27  \\ 
6162.173 & 1.899 & -0.09  \\ 
6163.754 & 2.521 & -1.29  \\ 
6166.440 & 2.521 & -1.14  \\  
6717.687 & 2.710 & -0.52  \\
\multicolumn{3}{c}{Sc {\sc ii}}  \\ 
5667.164 & 1.500 & -1.11$^a$ \\
5669.040 & 1.500 & -1.00$^a$  \\
5684.190 & 1.510 & -0.92$^a$ \\
6245.637 & 1.507 & -1.05    \\
\multicolumn{3}{c}{Ti {\sc i}}   \\
5689.487 & 2.297 & -0.47  \\
\multicolumn{3}{c}{Mn {\sc i}}  \\ 
6013.498 & 3.07  & -0.25$^a$ \\
6016.637 & 3.07  & -0.09$^a$ \\
6021.802 & 3.08  &  0.03$^a$ \\ 
\multicolumn{3}{c}{Ni {\sc i}} \\ 
6384.663 & 4.154 & -1.00  \\
6482.796 & 1.936 & -2.78  \\
\multicolumn{3}{c}{Ba {\sc ii}}  \\ 
5853.675 & 0.600 & -1.00$^a$  \\
6141.718 & 0.700 & ~0.00$^a$   \\
6496.896 & 0.600 & -0.38$^a$  \\
\multicolumn{3}{c}{Eu {\sc ii}}  \\ 
6645.110 & 1.380 & 0.204  \\
\hline
\multicolumn{3}{l}{$^a$: line for which HFS has been considered}\\
\end{tabular}
\label{t:lines}
\end{table}

\section{Other abundances}

Abundance derivation for other elements was also based on spectrum synthesis. 
Table~\ref{t:lines} lists the 
excitation potentials and oscillator strengths adopted for each atomic
line synthetized (C and N, for which we only studied the molecules CH, 
C$_2$ and CN are not included);
these line parameters are the same used throughout our BOCCE program. 
In summary, we have determined abundances of C, N, O, Na, Al, of the
$\alpha$-elements Mg, Si, Ca, Ti, of Sc, Mn, Ni, Ba, Eu.
Results for the individual stars
in the two clusters are presented in Table ~\ref{tab-ele}, while Table
~\ref{tab-mean} gives the average values.

We have determined the sensitivity of abundances to variations in atmospheric 
parameters using star 4510 in NGC~6253 (see Table~\ref{tab-sens}); given
the similarity of evolutionary status and metallicity, these
are valid also for the other stars in this cluster and in NGC~6791. 

\begin{table}
\centering
\caption{Sensitivity of derived abundances to variations
in atmospheric parameters.}
\begin{tabular}{lcccc}
\hline
Element    &$\Delta$ $T_{eff}$ 
           &$\Delta \log g$ 
	   &$\Delta$[X/H] 
	   &$\Delta v_t$\\
           &+100 K   
	   &+0.2 dex	 
	   &+0.1 dex 
	   &+0.1 km~s$^{-1}$\\
\hline
${\rm [C/Fe]}${\sc i}	 &-0.181 &~0.018 &-0.006 &~0.046\\
${\rm [N/Fe]}${\sc i}	 &-0.247 &~0.066 &-0.038 &~0.047\\
${\rm [O/Fe]}${\sc i}	 &-0.020 &-0.029 &-0.037 &~0.027\\
${\rm [Na/Fe]}${\sc i}   &~0.076 &-0.068 &-0.015 &~0.024\\
${\rm [Mg/Fe]}${\sc i}   &-0.023 &-0.029 &-0.011 &~0.023\\
${\rm [Al/Fe]}${\sc i}   &~0.048 &-0.025 &-0.029 &~0.012\\
${\rm [Si/Fe]}${\sc i}   &-0.093 &~0.007 &~0.000 &~0.036\\
${\rm [Ca/Fe]}${\sc i}   &~0.096 &-0.073 &-0.011 &~0.010\\
${\rm [Sc/Fe]}${\sc ii}  &~0.124 &-0.036 &-0.009 &-0.008\\
${\rm [Ti/Fe]}${\sc i}   &~0.126 &-0.032 &-0.018 &-0.025\\
${\rm [Mn/Fe]}${\sc i}   &~0.074 &-0.049 &~0.008 & 0.019\\
${\rm [Fe/H]} ${\sc i}   &~0.018 &~0.022 &~0.026 &-0.053\\
${\rm [Fe/H]} ${\sc ii}  &-0.146 &~0.124 &~0.044 &-0.036\\
${\rm [Ni/Fe]}${\sc i}   &-0.033 &~0.016 &~0.002 &-0.003\\
${\rm [Ba/Fe]}${\sc ii}  &~0.168 &-0.080 &~0.006 &-0.027\\
${\rm [Eu/Fe]}${\sc ii}  &~0.138 &-0.039 &-0.010 &-0.020\\
\hline
\end{tabular}
\label{tab-sens}
\end{table}

\subsection{NGC~6253}

\paragraph{C, N, O -}

For stars of this temperature and metallicity there is coupling between C and O
due to the formation of CO; furthermore, the O line used (6300.3 \AA)  is
contaminated by lines of the CN red system, meaning that we also need to
consider N abundance in our computations. The abundances of C, N and O were
derived following the classical procedure  first described in Lambert \& Ries
(1978, 1981). We used the version applied by Gratton \& Sneden (1990) to $\mu$
Leo, solving the full set of dissociation equations related to the coupling of
these elements including among others the CO and CN molecules.

Briefly, we obtained a first set of C and O abundances by assuming a guess for
the N abundance (actually, the solar-scaled [N/Fe] value); the best C and O
abundances were obtained from the intersection of the loci in the C-O plane
derived $via$ synthetic spectrum comparison with observed O and C features.

The oxygen abundances were derived from careful synthesis of the [O {\sc i}] 
6300.3~\AA\ line, the best   abundance indicator for red giants in the optical.
We did not need to correct for telluric components the spectra of the stars
observed with UVES, since we checked that the oxygen line was not affected by
them by comparing to the spectrum of  a rapidly rotating early type star. For
star 3595, observed with FEROS, the heliocentric correction is very different; 
the presence of contamination by telluric lines was detected and cleaned up by
using the same early type star spectrum. The contribution by  the Ni {\sc i}
line at 6300.35~\AA, whose effect is not negligible in moderately O-poor,
metal-rich stars in open clusters (see Carretta et al. 2005), was duly
considered adopting a solar [Ni/Fe] ratio: we included  the $\log gf$ for this
line by Johansson et al. (2003) in the list for synthetic spectra.

Very good estimates of the C abundances can be obtained from the spectral
synthesis of the  C$_2$ molecular features at 5086~\AA. However, only the
spectra obtained with FEROS contains this region; hence, the fitting was
possible only for star 3595.  All the stars in our sample have the same
evolutionary status and a quite similar appearance of their spectra, so that we
should be justified in adopting a value [C/Fe]$=-0.20$ dex also for the three
stars observed with UVES.   However, we confirmed the validity of this
assumption with synthesis of the CH band near 4300 \AA, present in all spectra.
With this C abundance, we derived from the forbidden [O {\sc i}] line trial
values for their O abundances.

With these preliminary values of the C, O abundances we derived abundances of N
from the comparison of synthetic spectra to a number of features  of the CN red
system, between 8158 and 8224~\AA\footnote{
This wavelength range was not observed with SARG, hence the N abundance for the
stars in NGC~6791, derived from weaker lines at shorter wavelengths, is more
uncertain}. 
We could match 15 to 24 such features,
depending on the star. The averaged N values, having an rms of 0.10 to 0.16
dex, are shown in Table 5 for each individual star. We find on average a small
depletion of N, with [N/Fe]$=-0.08$ (rms=0.03 dex, four stars). 

Finally, with these N values we reiterated the derivation of C
and O; however, results were found to be scarcely sensitive to the newly
derived N abundances, with respect to the initial solar-scaled value adopted.
In Table~\ref{tab-ele} we summarize the full set of C, N, O abundances obtained
with this procedure. 

\paragraph{The light elements Na and Al -}

Since the strong Na doublets at 5682-88 and 8183-8194~\AA\ are heavily saturated
and a reliable abundance is difficult to obtain from these features even with
spectrum synthesis, our Na values for NGC~6253 were derived from the weaker
doublet at 6154-60~\AA. The latter has the further advantage of  smaller
departures from the LTE assumption.

In Table~\ref{tab-ele} we list the [Na/Fe] ratios obtained for each star; the 
corrections for departures from LTE were found to be
negligible, following the prescription by Gratton et al. (1999). 
As found for the majority of open clusters studied so far, the average Na
in NGC~6253  is enhanced relative to the solar ratio, with no 
significant star-to-star
scatter (average [Na/Fe]=$+0.21$, rms=0.02 dex).

Abundance of Al were derived from spectrum synthesis of the doublet at
6696-98~\AA. We measured a [Al/Fe] slightly lower than solar-scaled 
(on average  [Al/Fe]$-0.08$, rms=0.12 dex).

\paragraph{$\alpha$-elements -}

We measured Mg abundances through spectrum synthesis of the Mg {\sc i} triplet
at 6318-19~\AA,  and of the line at 7691.5 \AA. The Mg lines are all strong and
this triplet, although largely used, is not a perfect abundance indicator since
damping is important and not well known\footnote{
Collisional damping used through this paper follows the precepts by
Barklem et al. (2000). However, Mg lines have a large effective number $n^*$ and
are then outside the validity range of these approximations. }.
We found a solar scaled ratio; on average [Mg/Fe]=$+0.01$ (rms=0.03) dex.
Although no significant star-to-star scatter is present, we attribute a larger
uncertainty to our measure for the reason given just above.

\begin{figure}
\centering
\includegraphics[bb=40 160 570 700, clip,scale=0.45]{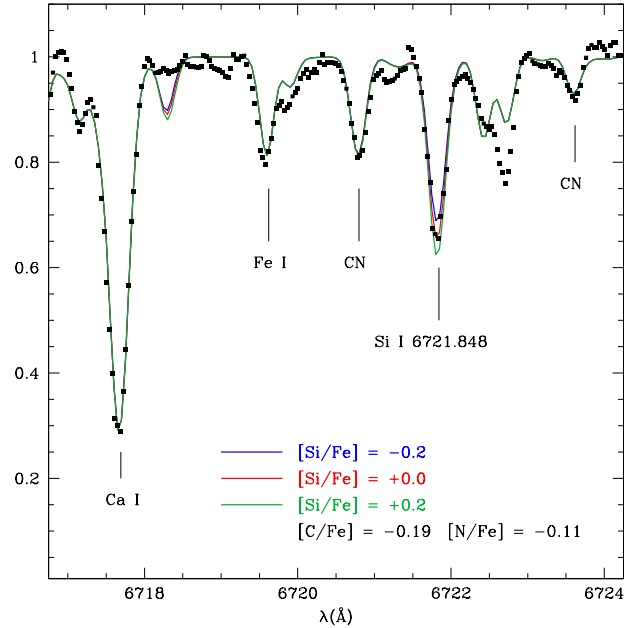}
\caption{Comparison of synthetic spectra with the Si {\sc i} 6721.85~\AA\ line
for star 2509 in NGC~6253. The observed spectrum is plotted as black dots and
the three synthetic spectra, obtained by changing the Si abundance in 0.2 dex
steps, are indicated by thin lines. The C and N abundances are those derived
for this star and nicely reproduce the nearby CN features, supporting the
reliability of [C/Fe] and [N/Fe] ratios obtained for this star.
}
\label{fig-si}
\end{figure}

The Si abundances are based on three lines (six for the FEROS spectrum), among
them the  clean Si {\sc i} 6721.85~\AA\ line that
falls in a region almost free from telluric
contamination and  where the spectra have a very high S/N. An example of the
fitting for this line in star 2509 is shown in Fig. \ref{fig-si}. A few other
features are also indicated; in particular, we can appreciate from this figure
how the derived C, N abundances reproduce fairly well the clean CN feature just
blueward of the Si {\sc i} line. We find
an average [Si/Fe]$+0.09$ (rms=0.06 dex).

We used spectrum synthesis of three or four moderately weak lines  of Ca {\sc i}
(at 6156, 6161, 6163 and 6166 ~\AA)  and checked them with the
6717~\AA\ line (which is blended with Fe {\sc i}, hence provides less reliable
abundances). They give the average value 
[Ca/Fe]=$-0.02$ (rms=0.12) dex.
We wish to note that our synthesis well reproduces the wings of
the Ca {\sc i} 6162.17 \AA \ line, a very good indicator of surface gravity
(see e.g., Mishenina et al. 2006, who used it in field RC stars, and literature
therein); this is an independent validation of the gravities we adopted.

Only one Ti {\sc i} line was available for synthesis, and only in star
3595; we derived the rather uncertain value of [Ti/Fe]$=-0.19$, that we retain
since this same line was measured in all four NGC~6791 stars.

The average of $\alpha$ elements is [$\alpha$/Fe]$=-0.03$ (rms=0.12) if we
consider all four species, and [$\alpha$/Fe]$=-0.03$ (rms=0.07) if we exclude Ti.
In both cases this value conflicts with the earlier claim of
$\alpha$ enhancement by  Twarog et al.
(2003) and confirms their later conclusion (Anthony-Twarog et al.
2007; see also Sestito et al. 2007 for a
very similar result). 

\paragraph{Iron peak elements -} 

We measured the abundances of Sc, Mn and Ni.  Of the four Sc {\sc ii} lines, only
one was available in all stars; for the three others we considered the hyperfine
structure (HFS, see Gratton et al. 2003 for references).  The Sc abundance was
found remarkably constant and we derived the average abundance
[Sc/Fe]$=-0.06$ (rms=0.01) dex.

The Mn abundance was derived from the three saturated lines at 6013-6021 \AA,
admittedly not very sensible to abundance variations. We took HFS into
consideration; the three lines gave reasonably consistent values and we found an
average [Mn/Fe]$=+0.04$ (rms=0.07) dex.

The Ni abundance is based on two lines, and the average value is 
[Ni/Fe]$=-0.05$, without scatter.

All three Fe-peak elements have solar scaled ratios; this is expected for Sc and
Ni. The "right" Mn abundance is more difficult to tell, since it appears
decoupled from the iron one for most metallicities (Gratton et al. 2004). 
Furthermore, there are very few measures for open clusters, and none for so
metal-rich ones (see e.g., De Silva et al. 2007).

\paragraph{Neutron capture elements -} 

The Ba abundance was derived from two or three strong lines, for which HFS 
(from Prochaska et al. 2000) was
considered; the final average value is [Ba/Fe]{\sc ii}$=+0.16$ (rms=0.04).  This
element has been measured for many OCs, finding a very scattered distribution,
with abundances varying by more than 1 dex (e.g., De Silva et al. 2007). 

Finally, we measured one Eu line, finding the remarkably constant
value [Eu/Fe]{\sc ii}$=+0.03$.

\begin{table*}
\centering
\caption{
Elemental ratios for the individual stars in NGC~6253 and NGC~6791.
We indicate the solar reference values (Col. 2) and, for each star, the number of
lines and the abundance ratio and the rms.
}
\begin{tabular}{lrrrrrrrrrrrrrr}\\
\hline
element           & Sun       &nr &[X/Fe] &rms  &nr &[X/Fe] &rms &nr &[X/Fe] &rms  &nr &[X/Fe] &rms    \\ 
\hline
& &\multicolumn{3}{c}{NGC~6253-2509}     
 &\multicolumn{3}{c}{NGC~6253-2885}	 
 &\multicolumn{3}{c}{NGC~6253-4510}	 
 &\multicolumn{3}{c}{NGC~6253-3595} \\   
C {\sc i}  &8.52 &10 &-0.19 &0.11  &10 &-0.15 &0.18  &11 &-0.13 &0.14 &6  &-0.10  &0.11 \\
N {\sc i}  &7.92 &15 &-0.11 &0.12  &22 &-0.18 &0.11  &24 &-0.13 &0.16 &20 &-0.12  &0.10 \\
O  {\sc i} &8.79 & 1 &-0.15 &	   & 1 &-0.25 &	     &1  &-0.20 &     & 1 &-0.12  &     \\
Na {\sc i} &6.21 & 2 & 0.21 &0.04  & 2 & 0.23 &0.07  &2  & 0.18 &0.00 & 2 & 0.23  &0.07 \\
Mg {\sc i} &7.43 & 4 & 0.00 &0.04  & 4 & 0.02 &0.09  &3  & 0.05 &0.04 & 4 &-0.03  &0.01 \\
Al {\sc i} &6.23 & 2 &-0.12 &0.01  & 2 &-0.19 &0.07  &2  &-0.09 &0.01 & 2 & 0.10  &0.01 \\
Si {\sc i} &7.53 & 3 & 0.01 &0.19  & 3 & 0.07 &0.21  &3  & 0.14 &0.20 & 6 & 0.14  &0.13 \\
Ca {\sc i} &6.27 & 4 & 0.00 &0.10  & 4 &-0.06 &0.10  &4  &-0.16 &0.13 & 3 & 0.13  &0.15 \\  
Sc {\sc ii}&3.13 & 1 &-0.06 &	   & 1 &-0.06 &	     &1  &-0.06 &     & 3 &-0.05  &0.10 \\
Ti {\sc i} &5.00 &   &	    &	   &   &      &	     &   &      &     & 1 &-0.19: &     \\
Mn {\sc i} &5.34 & 3 & 0.03 &0.10  & 3 & 0.10 &0.09  &3  & 0.07 &0.08 & 2 &-0.05  &0.14 \\
Ni {\sc i} &6.28 & 2 &-0.05 &0.27  & 2 &-0.05 &0.27  &2  &-0.05 &0.27 & 2 &-0.05  &0.02 \\
Ba {\sc ii}&2.22 & 2 & 0.13 &0.14  & 2 & 0.18 &0.21  &2  & 0.13 &0.14 & 3 & 0.20  &0.16 \\
Eu {\sc ii}&0.55 & 1 & 0.03 &	   & 1 & 0.03 &	     & 1 & 0.03 &     & 1 & 0.03  &     \\
\hline
&&\multicolumn{3}{c}{NGC~6791-2014} 
 &\multicolumn{3}{c}{NGC~6791-3009}
 &\multicolumn{3}{c}{NGC~6791-3019}
 &\multicolumn{3}{c}{NGC~6791-SE49}   \\
C {\sc i}  &8.52 & 1 &-0.18 &     & 1  &-0.12 &     & 1 & -0.23  &     & 1 &-0.38 &     \\
O {\sc i}  &8.79 & 1 &-0.35 &     & 1  &-0.20 &	 & 1 & -0.35  &     & 1 &-0.35 &        \\
Na {\sc i} &6.21 & 1 & 0.28 &     &    &      &	 &   &        &     & 1 &-0.02 &        \\
Mg {\sc i} &7.43 & 3 & 0.24 &0.09 & 3  & 0.24 &0.14 & 3 &  0.18  &0.23 & 3 & 0.14 &0.15 \\
Al {\sc i} &6.23 & 2 &-0.24 &0.01 & 2  &-0.17 &0.01 & 2 & -0.32  &0.01 & 2 &-0.12 &0.01 \\
Si {\sc i} &7.53 & 4 &-0.06 &0.17 & 4  & 0.12 &0.17 & 3 & -0.11  &0.10 & 4 & 0.02 &0.11 \\
Ca {\sc i} &6.27 & 4 &-0.16 &0.08 & 1  &-0.06 &	    & 1 & -0.26  &     & 5 &-0.13 &0.19	\\
Sc {\sc ii}&3.13 & 3 &-0.23 &0.14 & 4  &-0.08 &0.13 & 3 & -0.10  &0.09 & 3 &-0.10 &0.01 \\
Ti {\sc i} &5.00 & 1 & 0.01 &     & 1  &-0.04 &	    & 1 &  0.16  &     & 1 & 0.01 &     \\
Mn {\sc i} &5.34 & 3 & 0.13 &0.16 & 3  & 0.10 &0.22 & 3 &  0.10  &0.10 & 3 &-0.03 &0.03 \\
Ni {\sc i} &6.28 & 2 & 0.00 &0.23 & 2  &-0.03 &0.33 & 2 & -0.15  &0.13 & 2 &-0.08 &0.33 \\
Ba {\sc ii}&2.22 & 3 & 0.43 &0.10 & 3  & 0.31 &0.03 & 3 &  0.21  &0.20 & 3 & 0.15 &0.18 \\
Eu {\sc ii}&0.55 & 1 &-0.17 &     & 1  &-0.17 &	    &   &        &     &   &      &     \\
\hline
\end{tabular}
\label{tab-ele}
\end{table*}

\subsection{NGC~6791}

In G06 we determined Fe, C, O (and N) abundances in  NGC~6791. Here we applied
to our spectra  the same technique described above for NGC~6253 but given the
different instrument, wavelength coverage and S/N not all lines synthesized for
NGC~6253 were available for NGC~6791 too.

Unfortunately the region near 6150-6170 \AA~ falls near the separation between
the two chips in the SARG mosaic with our setting and it was either missing or
badly extracted in several spectra, depending on the exact location of the stars
along the slit.  For this reason we were able to reliably determine the Na
abundance only for two stars (2014 and SE49) and using only the 6154 \AA ~line.
At variance with NGC~6253, where the Na  abundance shows no scatter, we found
two quite differing values, giving an average ${\rm [Na/Fe]}=+0.13$ (rms=0.21)
dex. 

Also the Ca abundance was derived using the four lines between 6156.10 and 
6166.44 \AA \ only for the same two stars; for the two others we had to use  
only the 6717.69 \AA \ line; there is no significant scatter for
the Ca abundance and the average value is  ${\rm [Ca/Fe]}=-0.15$ (rms=0.08)
dex. Mg was measured only from the triplet near 6319 \AA ~and the abundance 
ratio found is enhanced with respect to the solar ratio
(${\rm [Mg/Fe]}=+0.20$, rms=0.05).
The Ti line chosen was instead visible in all
four stars and the average value is ${\rm [Ti/Fe]}=+0.03$ (rms=0.09).
In the case of NGC~6791 ${\rm [\alpha/Fe]}=+0.02$ (rms=0.14) dex.

Individual results for the four stars and average cluster values 
are presented in Tables ~\ref{tab-ele} and \ref{tab-mean} respectively. Here 
we also
show the individual O and C abundances, while in  G06 we only presented the
averages for the four stars. Note that the present values slightly differ
from the ones given in G06, since we used there different solar reference
values. The present ones are as in Gratton et al. (2003) and Carretta et
al. (2005) and will consistently be used for all the BOCCE clusters.

\begin{table}
\centering
\caption{Average elemental ratios for NGC~6253 and NGC~6791. The last column
gives the corresponding  values for $\mu$ Leo, taken from Gratton \& Sneden
(1990), but   modified from the original ones to take into account the present
reference solar values.
}
\begin{tabular}{lrrrrr}\\
\hline
element &\multicolumn{2}{c}{NGC~6253} &\multicolumn{2}{c}{NGC~6791} &$\mu$ Leo\\ 
& mean &rms  & mean &rms & \\               
\hline
${\rm [Fe/H]}$    & $+0.46$ &0.03   & $+0.47$ &0.07 & $+0.46$ \\ 
${\rm [C/Fe]}$    & $-0.14$ &0.04   & $-0.23$ &0.11 & $-0.13$ \\ 
${\rm [N/Fe]}$    & $-0.14$ &0.03   & $-0.26$ &     & $+0.32$ \\
${\rm [O/Fe]}$	  & $-0.18$ &0.06   & $-0.31$ &0.08 & $-0.13$ \\ 
${\rm [Na/Fe]}$   & $+0.21$ &0.02   & $+0.13$ &0.21 & $+0.56$ \\
${\rm [Mg/Fe]}$   & $+0.01$ &0.03   & $+0.20$ &0.05 & $-0.07$ \\ 
${\rm [Al/Fe]}$   & $-0.08$ &0.12   & $-0.21$ &0.09 & $+0.30$ \\ 
${\rm [Si/Fe]}$   & $+0.09$ &0.06   & $-0.01$ &0.10 & $+0.08$ \\ 
${\rm [Ca/Fe]}$   & $-0.02$ &0.12   & $-0.15$ &0.08 & $-0.12$ \\ 
${\rm [Sc/Fe]}$   & $-0.06$ &0.01   & $-0.13$ &0.07 & $-0.06$ \\
${\rm [Ti/Fe]}$   & $-0.19$:&	    & $+0.03$ &0.09 & $-0.05$ \\ 
${\rm [Mn/Fe]}$   & $+0.04$ &0.07   & $+0.08$ &0.07 & $-0.08$ \\ 
${\rm [Ni/Fe]}$   & $-0.05$ &0.00   & $-0.07$ &0.07 & $+0.06$ \\
${\rm [Ba/Fe]}$   & $+0.16$ &0.04   & $+0.28$ &0.12 &         \\
${\rm [Eu/Fe]}$   & $+0.03$ &0.00   & $-0.17$ &0.00 &         \\
\hline
\end{tabular}
\label{tab-mean}
\end{table}

\section{Literature studies}

\subsection{Other analyses of the two clusters}

We will restrict our comparison to papers based on high resolution spectroscopic
studies in the case of NGC~6791, but consider also intermediate-band photometry
for NGC~6253, for which the literature is far less copious.

Initial studies of NGC~6253 concentrated on  broad-band photometry 
(Bragaglia et al. 1997; Sagar et al. 2001), in one case coupled with 
integrated spectra  (Piatti et al. 1998). These papers agreed on a
rather old age (from 2.5 to about 5 Gyr) and a high metallicity ([Fe/H]=+0.2
to +0.4).

As part of their series on Str\"omgren photometry of open clusters, Twarog et
al. (2003) presented a study based on $uvbyCa{\rm H}\beta$ photometry (and a
comparison to literature broad-band data).  NGC~6253 is outside the calibrations
available at the time of the $\delta m_1$ and $\delta hk$ indices. With some
extrapolations, they derived formal values of [Fe/H]  of +0.7 to +0.9 dex, but
their best fit with isochrones led to [Fe/H] closer to +0.4, a significant
$\alpha$-enhancement, and an age between 2.5 and 3.5 Gyr. In Anthony-Twarog et
al. (2007) they revised these results using an updated version of the
colour-metallicity relations for Str\"omgren photometry, valid in the
super-metal rich regime, and favoured instead solar scaled $\alpha$-elements, a
reddening $E(B-V)=0.16\pm0.025$, an age of about 3 Gyr, and
[Fe/H]$=+0.58\pm0.04$.
Following the reasoning in G06 (sect. 3.3 and Table 4), had we adopted their
lower reddening instead of the value 0.23, we would have obtained a metallicity about
0.13 dex higher (the decrease in temperature also implies a decrease in
microturbulent velocity and [Fe/H] is much more sensible to variations in the
latter at these metallicities and temperatures). An important feature of their
reanalysis of NGC~6253 is that the similarity of results from the $\delta m_1$
index (that is dominated by CN features) and the $\delta hk$ index (dominated by
Ca) should indicate that [C/Ca]$\sim$0 for this cluster (we actually find a
value of 0.1 dex).  This is at variance with NGC~6791, where the two indices give
discrepant results; Anthony-Twarog et al. (2007) explain this with C (and N, but
not Ca) being underabundant in NGC~6791 with respect to solar. While C and N
have less undersolar ratios in NGC~6253 than in NGC~6791 we do not know if this
is enough to explain the different behavior of the metallicity indices in the
two clusters. 

We presented a preliminary metallicity for NGC~6253 (Carretta et al. 2000) based
on two objects, the red clump stars 2508, later reobserved, and 2971. The
spectra were obtained on July 1998 with EMMI on the ESO New Technology Telescope
as a backup in poor weather conditions, at R=28000, degraded to about
R$\simeq$15000 to improve the rather low S/N. From equivalent widths (EWs)
analysis we found an average  [Fe/H]=$+0.36\pm0.20$ and concluded that the value
could be taken as indicative of a metal abundance larger than solar, but with a
large error attached. Further spectra at higher resolution and S/N were deemed
necessary to conclusively define the cluster metallicity; the present work
supersedes the old analysis.

The only other detailed analysis of high resolution spectra of NGC~6253 is by
Sestito et al. (2007), who analyzed 5 stars observed with UVES/FLAMES, in
different evolutionary phases (RC, RGB, SGB and Turn-Off). There are no stars  
in common with our set.
Since they were mostly dealing with warmer stars than the ones studied here, 
they used EWs, only using synthetic spectra to check their results. For the
elements in common, they derived
[Fe/H]=$+0.39$, [Si/Fe]=$+0.02$, [Ca/Fe]=$-0.04$, [Ti/Fe]=$-0.01$,
[Mg/Fe]=$+0.30$, [Ni/Fe]=$+0.08$, [Ba/Fe]=$+0.23$, and [Na/Fe]=$+0.07$. 
These values are in agreement with ours, with the exception of Ti (but our
abundance is based on only one line in one star), Mg (that both papers consider
among the less reliable  derivations), and Na (for which two different
corrections for NLTE were applied; actually their average LTE value is +0.20,
similar to ours).

The literature on NGC~6791 is far more abundant because of its unique
combination of very old age and high metallicity, that raised considerable
interest. 
After the work by Peterson \& Green (1998) who analyzed a single red horizontal
branch star, this old, metal rich cluster has been the subject of three recent
independent papers based on high resolution spectra. Our own (G06),  Carraro et
al. (2006) who measured [Fe/H] and elemental abundances for a sample of ten red
giants observed with HYDRA at R=17000 finding [Fe/H]=+0.39, and Origlia et al.
(2006) who used the near IR spectra at R=25000 of six cold giants obtained with
NIRSPEC at R=25000 to determine[Fe/H]=+0.35 and several elemental abundances.
Finally, in a very recent paper Anthony-Twarog et al. (2007) derived
[Fe/H]=+0.45, based on the version of the colour-metallicity relations for
Str\"omgren photometry valid for very high metallicity stars.

We will compare our findings to the spectroscopic works by Peterson \&
Green (1998), Carraro et al. (2006) and Origlia et al. (2006). 
Table~\ref{tab-conf} shows the abundances for the species in common; errorbars
on these elemental ratios are of the order of 0.1 or 0.2 dex, depending on the 
species, in all papers. The agreement between different analyses is far from
satisfactory. Iron shows formally the best accord since all measures agree
within the quoted errorbars. Our measures are in good agreement with the others'
for some elements (e.g., for Mg, Si, Ti, Ni) while there are significant
differences for other elements (e.g., for O, Na, Ca, and most notably N).  We
have no real explanation for this. 
A real comparison is difficult even if all investigators
employed spectrum synthesis
to analyze the data, because the studies use different stars
and systematics can be hard to assess. Peterson \& Green (1998) analyzed a
single red HB star, i.e., a star much hotter (about 7300 K) than
everyone else; Origlia et al. (2006) studied cold M giants (3600-4000 K) in a
completely different wavelength range (the near IR); finally, Carraro et al.
(2006) chose stars similar to ours (RC and RGB stars of magnitude comparable to
the RC). 

The latter study has also one star in common with us (their 8082, our SE49) and
we may attempt some direct comparison. 
Both studies obtained for this star spectra of similar S/N (about 40), and  measured a
similar radial velocity ($-46.18$ km~s$^{-1}$ for them and $-45.63$ km~s$^{-1}$
for us) but our spectra have a resolution of about 30000 while theirs of about
17000.  They assumed a distance modulus
$(m-M)_0=12.79$, with $E(B-V)=0.09$ (from Stetson et al. 2003) i.e., 
$(m-M)_V=13.07$, and derived temperatures from $B-V$ and $V-I$ colours using
three different relations. We adopted $(m-M)_V=13.45$, with $E(B-V)=0.15$ (as an
average of literature determinations) and derived the temperature from $V-K$ and
the Alonso et al. (1999) relations. The adopted atmospheric parameters for
temperature, gravity and microturbulent velocity are quite different: 
$T_{eff}=4410$, $\log g=2.81$, $v_t=1.00$ for Carraro et al. (2006) and
$T_{eff}=4512$, $\log g=2.32$, $v_t=1.05$ for G06. 

The difference in temperature is mainly due to the different reddening values
adopted; had Carraro et al. (2006) adopted the same higher reddening we used,
they would have obtained temperatures similar to ours. The differing gravity is
only partly due to the combination of lower temperature and shorter distance
modulus. The large residual difference would be explained had they used a  mass
of about 2.5 M$_\odot$, compared to the 0.9 M$_\odot$ we adopted; they do not
give explicitly the value, but this is what we understand from the spectral
type and the Straizys \& Kuriliene (1981) work referenced in their paper. This
large mass is in contrast with the old age of the cluster.

Is there any way of independently checking the soundness of the
assumptions on distance and reddening? We have not yet studied NGC~6791
using the synthetic colour-magnitude diagrams technique adopted for the
BOCCE sample; however, we note that our values are
close to the ones found by King et al. (2005) on HST/ACS photometric data
and to the ones derived by Anthony-Twarog et al. (2007)
on Str\"omgren data. The latter also provide (their sect. 6.3) an
interesting discussion on
the validity of different comparisons between the photometric data 
and isochrones and on the inconsistency of a reddening as low as $E(B-V)=0.09$.
Carraro et al. (2006) provide a new
determination of age, distance and reddening using their derived metallicity and
two isochrone sets. From the Girardi et al. (2000) ones they confirm the
reddening but find a best fit with $(m-M)_V=13.35$ or 13.45 depending on the
age; from the Y$^2$ ones (Yi et al. 2001) they find best fit values of
$(m-M)_V=13.35$ and $E(B-V)=0.13$. However, they do not discuss the impact of
these new parameters on their metallicity derivation.
G06 discussed (their Sect. 3.3) the adopted temperature and reddening, finding
that strongly differing values would result in gravities implying
absolute magnitudes of the RC stars incompatible with what is expected from
evolutionary models (i.e., $M_V\sim1.2$, Girardi \& Salaris 2001).
We note also that with our adopted gravities we were able to reproduce very
well the wings of the 6162.17 \AA \ Ca {\sc i} line  which,   as said in Sect.
4.1, is a good indicator of surface gravity.

The difference in adopted atmospheric parameters is not the main cause of
the (small) difference in metallicity, as can be seen from the sensitivity
of [Fe/H] on parameters variation (Table~6). Our analysis and the one by
Carraro et al. (2006) use different
synthesis codes and line lists ad cannot be readily compared. However, we
note that the not ideal resolution of the HYDRA spectra has
a direct impact on
continuum tracing.  From their fig. 4, the continuum level may have been
underestimated by about 2\%, implying a [Fe/H] lower by  0.1 dex.
This (systematic) error of about 0.1 dex is within the uncertainties that
can be reached with the quality of the HYDRA spectra for these very line-rich
stars. 
The uncertainty
in continuum tracing is lower in our spectra because of resolution and
S/N (R$\sim30000$ versus 
$\sim17000$  and S/N higher by about a factor 1.5).  
We conclude that the difference of 0.08 dex in the mean cluster
metallicity is well within of the (combined) uncertainties of the two
derivations.

Finally, the difference in mean cluster metallicity between our work
and Carraro et al. (2006) is small and has no impact on the science discussed
in the two papers.

\begin{table}
\centering
\caption{NGC~6791: comparison of abundances between our work and 
literature values (Peterson \& Green 1998 - PG98, Carraro et al. 2006 - C+06,
Origlia et al. 2006 = O+06).}
\begin{tabular}{lcccc}
\hline
Element & Present work & PG98 & C+06 & O+06 \\
\hline 
${\rm [Fe/H]}$  & $+0.47$ & $+0.4$ & $+$0.39 & $+$0.35\\ 
${\rm [C/Fe]}$  & $-0.23$ & $+0.0$ & $ $     & $-$0.35\\ 
${\rm [N/Fe]}$  & $-0.26$ & $+0.5$ & $ $     & $ $    \\
${\rm [O/Fe]}$	& $-0.31$ & $+0.0$ & $ $     & $-$0.07\\ 
${\rm [Na/Fe]}$ & $+0.13$ & $+0.4$ & $ $     & $ $    \\
${\rm [Mg/Fe]}$ & $+0.20$ & $+0.2$ & $ $     & $-$0.03\\ 
${\rm [Al/Fe]}$ & $-0.21$ & $+0.0$ & $-$0.15 & $+$0.05\\ 
${\rm [Si/Fe]}$ & $-0.01$ & $+0.2$ & $+$0.02 & $+$0.02\\ 
${\rm [Ca/Fe]}$ & $-0.15$ & $+0.0$ & $-$0.03 & $+$0.05\\ 
${\rm [Sc/Fe]}$ & $-0.13$ & $+0.0$ &         &        \\
${\rm [Ti/Fe]}$ & $+0.03$ & $+0.0$ & $-$0.02 & $+$0.03\\ 
${\rm [Ni/Fe]}$ & $-0.07$ & $+0.0$ & $-$0.01 &        \\
${\rm [Ba/Fe]}$ & $+0.28$ & $+0.0$ & $-$0.13 &        \\
${\rm [Eu/Fe]}$ & $-0.17$ & $+0.1$ &         &        \\
\hline
\end{tabular}
\label{tab-conf}
\end{table}

\begin{figure*}
\centering
\includegraphics[bb=20 170 585 695, clip, scale=0.9]{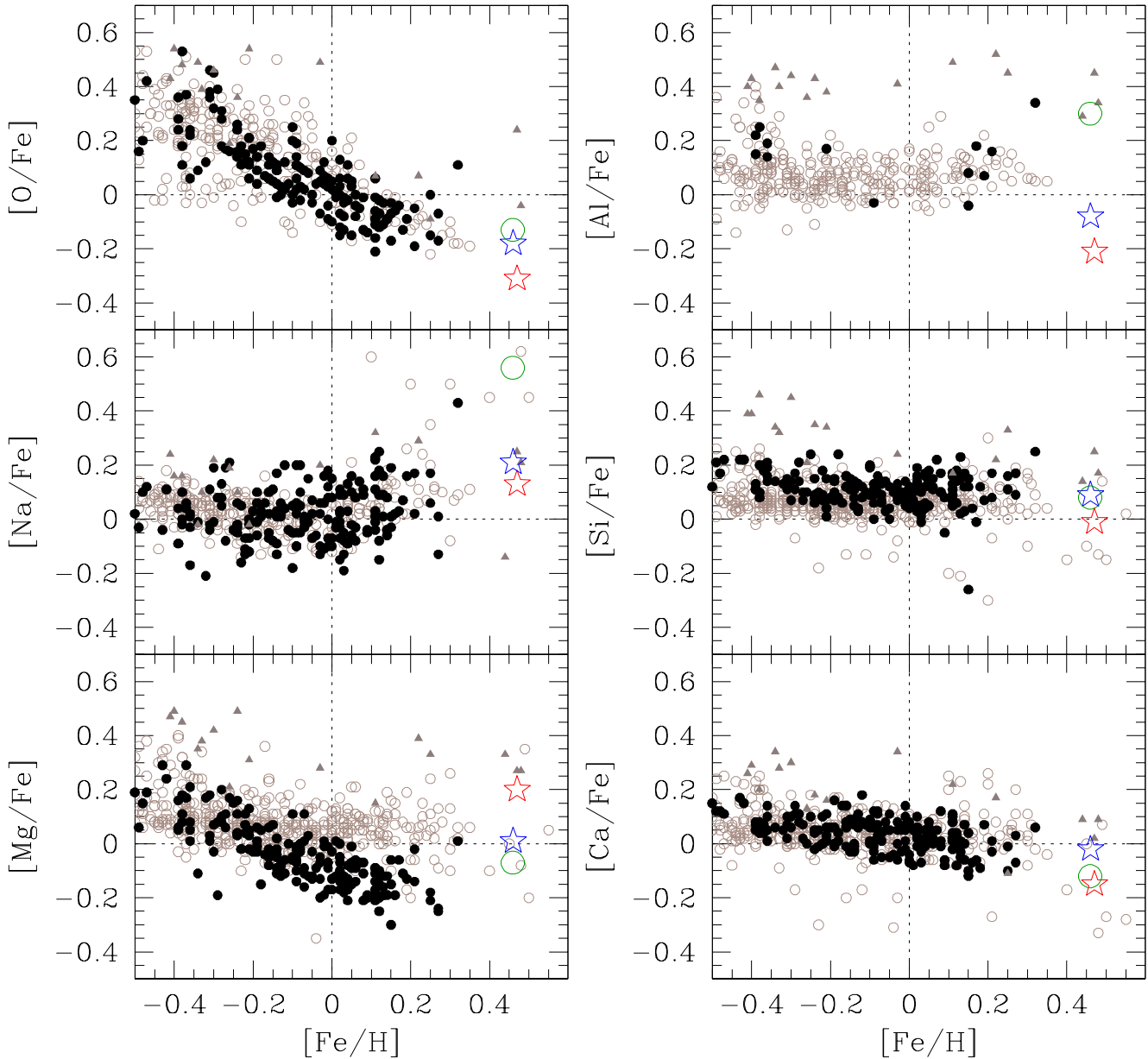}
\caption{Plot of the abundances (with respect to Fe) for the  light elements O,
Na, Mg, Al, Si, Ca $versus$ [Fe/H]  for NGC~6253, NGC~6791  and $\mu$ Leo (open
stars, blue and red respectively and open circle), compared to literature samples. We
plot with (grey) open circles the disk dwarf stars from Soubiran \& Girard
(2005), Castro et al. (1997) and Pompeia et al. (2003). Filled circles are giant
disk stars in Mishenina et al. (2006) and Fulbright et al. (2007); filled
triangles are giant bulge stars also from Fulbright et al. Note that the only
samples extending to the high metallicity of NGC~6253 and NGC~6791 are the one
by Castro et al. (they do not provide O and Al abundances) and the bulge one by
Fulbright et al.}
\label{fig-6el}
\end{figure*}
\begin{figure*}
\centering
\includegraphics[bb=20 170 585 695, clip, scale=0.9]{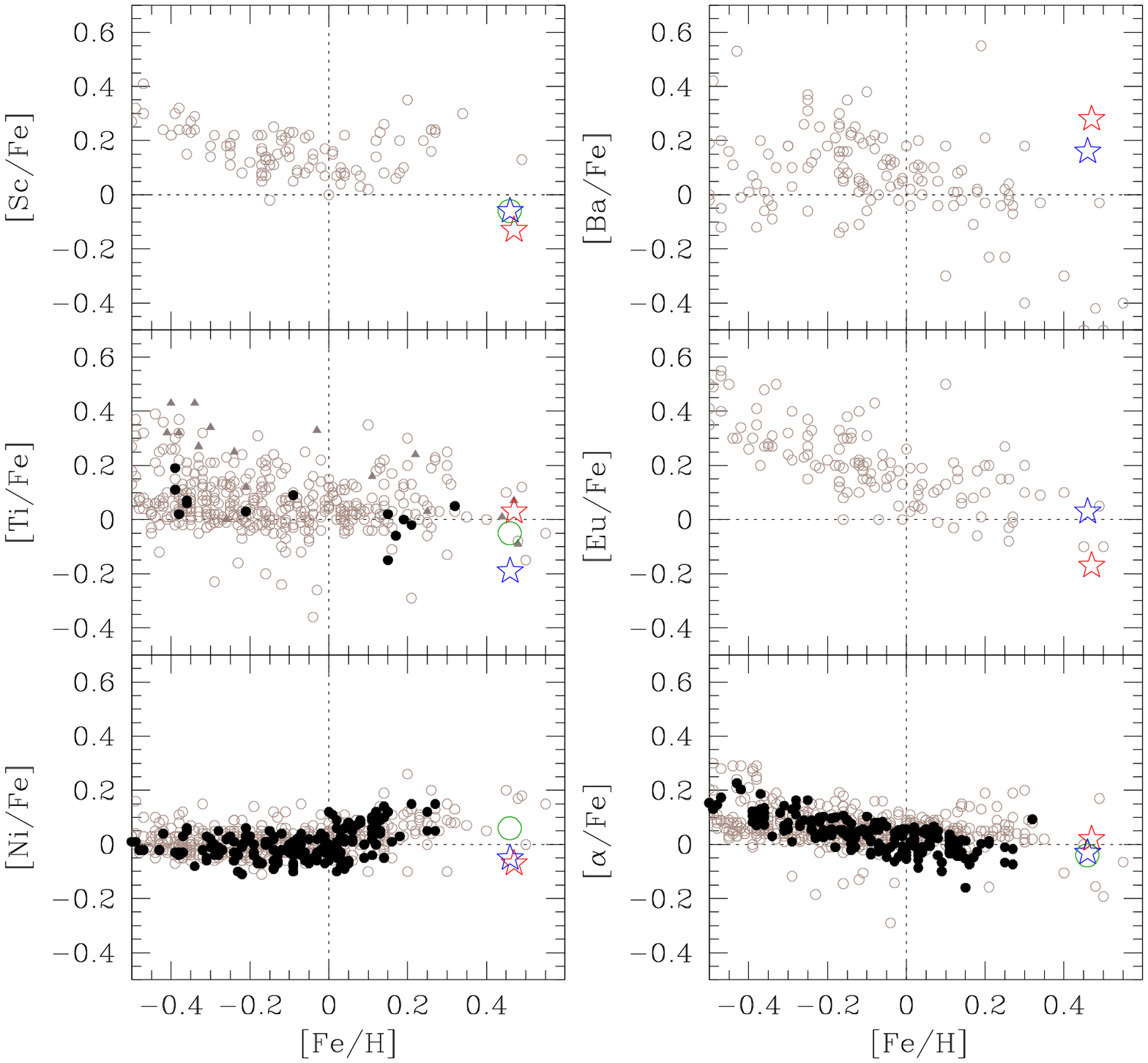}
\caption{Plot of the abundances of Sc, Ti, Ni, Ba, Eu and $\alpha$ elements
 $versus$ [Fe/H]  for NGC~6253, NGC~6791  and $\mu$ Leo  compared to
literature samples (symbols are as in the preceding Figure). Data for Sc 
and for Zr, Ba, Eu come
directly from the paper by  Allende Prieto et al. (2004), not rehomogeneized by
Soubiran \& Girard (2005); for the three latter, also data from Castro et al.
(1997) and Pompeia et al. (2003) were plotted.}
\label{fig-5el}
\end{figure*}

\subsection{Literature on high metallicity stars}

Are NGC~6253 and NGC~6791 rare and unusual ? The answer is important if we want
to gain more insight into their origin and  to use them to understand the
formation, the early enrichment and the metallicity distribution  of the disk.

These two OCs are so metal rich that it is not easy to find suitable
comparison samples in the literature, specially of giants,  to see if the
derived abundances follow the "normal" trends generally found
for disk objects. 
We consider here various possible comparisons: 
{\em i)} the very metal-rich giant star $\mu$ Leo, already
used in G06 to confirm the validity of our assumptions and method, with abundances taken
from Gratton \& Sneden (1990) but brought to our same solar reference scale  (see
Table~\ref{tab-mean}); {\em ii)}  the sample of thin and thick disk dwarfs in
Soubiran \& Girard (2005), who put together and tried to homogenize  a large
number of studies\footnote{
Allende Prieto et al. (2004), Bensby et al. (2003, 2004), Chen et al. (2000),
Edvardsson et al. (1993), Fulbright (2000), Gratton et al. (2003), Mishenina et
al. (2004), Nissen \& Schuster (1997), Prochaska et al. (2000), Reddy et al.
(2003).
}; 
{\em iii)}  the metal-rich local dwarfs by Castro et al.
(1997) and Pompeia et al. (2003), who selected stars of supposed inner  disk or
bulge origin on the basis of kinematics; {\em iv)}  the field disk RC stars
studied by Mishenina et al. (2006); {\em  v)}  the disk and the bulge
giants in Fulbright et al. (2007). As usually done when such comparisons are
presented, no effort was taken to  study or correct for systematics between
all samples;  we only note, for example, that $\mu$ Leo has [Fe/H]=+0.32 in
Fulbright et al. (2007) and +0.46 in our scale.

Fig.~\ref{fig-6el} shows the run of O, Na, Mg, Al, Si and Ca with metallicity
for all stars in these samples (open symbols are for dwarfs and filled symbols
for giants); the two OCs plus $\mu$ Leo are shown with larger symbols. The
only two disk samples that contain stars as metal-rich as NGC~6253 and NGC~6791
are the ones by Castro et al. (1997) and Pompeia et al. (2003), so we miss a
direct comparison for O and Al.

Taking into account that these ratios have an errorbar of about 0.1 dex, 
NGC~6253 and NGC~6791  display very similar abundance ratios, as seen
also from Table~\ref{tab-mean}, with the exception of Mg and Ti  (note however
that the Mg abundances have a larger errorbar attached -see  Sect. 4.1- and that
Ti is measured from one single line, and not in all stars).
Both clusters generally follow the  trend of the disk samples or its reasonable
extrapolation. The Mg abundance in NGC~6791 seems more similar to 
that of the bulge giants in Fulbright et al. (2007)  
but, given also that the distributions for disk
and giant stars for Mg differ while they do not for the other elements,
we do not consider this as
a strong indication that this cluster shows signature of a bulge-like origin.
Carraro et al. (2006) suggested that NGC~6791 is either of
extragalactic origin or originated in the inner disk region, near the bulge.
While the first hypothesis is completely unsupported by the abundance ratios
(ours or anyone else's), the second one has received a confirmation by a new
determination of the cluster absolute motion and orbit (Bedin et al. 2006).
Clearly, NGC~6791 would require new theoretical studies to explain the very
high metallicity  for its age and to better understand its origin;  however,
these studies would have to take into account that NGC~6791 is not perhaps an
exceptional object since there are at least two old clusters with similar high
metallicity and abundance ratios. For a deeper insight on these clusters'
properties and origin, new, higher quality spectra 
would also be useful: not only the abundances would stand on firmer grounds, but
new elements could be examined, e.g., adding more $s$- and $r$-process elements,
that could be used to shed light on the enrichment path and discriminate 
among different origins.
To this end we would benefit of a larger wavelength coverage, Especially in the
blue,  higher S/N and much higher resolution  than obtained by all the presently
existing studies.

\section{Summary}

With this paper we add two clusters to the very metal-rich end of the BOCCE
sample, that now comprises objects from [Fe/H]$=-0.2$ to more than +0.4 dex. The
fact that we concentrate on RC stars in all clusters helps to maintain a better
homogeneity in our results, even if different spectrograph, or  different
analysis methods (EWs or synthetic spectra) are used. Furthermore, we use the
same line lists, $gf$ values, model atmospheres, reference solar values and we always
check results of the EW analysis with synthesis, at least for Fe.  

We have analyzed high resolution spectra of four RC stars in NGC~6253  and
derived, using spectrum synthesis, their metallicity; the cluster has
[Fe/H]$=+0.46 \pm 0.02 \pm 0.08$ dex (internal and systematic error). A fifth
star is possibly a binary with the spectrum likely contaminated by
the secondary. For this reason, it was not taken into account in our results. We
also derived abundances for the light elements C, N, O, for Al and Na, for the
$\alpha$-elements  Mg, Si, Ca and Ti, for the iron-peak elements Sc, Mn, and Ni,
and for the neutron-capture elements Ba and Eu.

On the basis of the present analysis, NGC~6253 turns out to be as metal-rich as
NGC~6791, up to now the only cluster for which a metallicity well in excess of
the Hyades' had been convincingly derived, both on the basis of  photometric
data (e.g., Chaboyer et al. 1999) and of spectroscopic abundance analysis 
(Peterson \& Green 1998, Carraro et al. 2006, Origlia et al. 2006,  and G06). 

Our metallicity determination for NGC~6253 is in agreement with the  photometric
derivations (e.g., Bragaglia \& Tosi 2006, Anthony-Twarog et al.  2007). 
Also the comparison with Sestito et al. (2007) is favourable. We
have measured Mg, Si, Ca (and Ti)  abundances and the resulting [$\alpha$/Fe] is
solar, as also found recently by Sestito et al. (2007) and  Anthony-Twarog et
al. (2007).

We have also completed our analysis of  NGC~6791, deriving abundances for the
same elements as in NGC~6253 for four RC stars. The two clusters show generally
similar abundance ratios; an exception is Mg, which is overabundant in NGC~6791
-and more similar to values measured in bulge giants- but not in NGC~6253. 

In NGC~6791 the [$\alpha$/Fe] value is solar, confirming earlier results by
Carraro et al. (2006) and   Origlia et al. (2006) and only slightly differing
with Peterson \& Green (1998) who obtained  [$\alpha$/Fe]=+0.10. The agreement
appears quite good, but could be completely accidental because of the different
data sets and analysis methods.

The two clusters seem to follow rather well the trends of elemental ratios for
metal-rich stars, as derived from the comparison to different literature
samples, both for field dwarfs and giants and for bulge giants.  They generally
appear to conform to the disk stars run of elemental ratios with [Fe/H], with
the possible exception of Mg for NGC~6791.

\begin{acknowledgements}
This paper is part of the BOCCE project and we are grateful to M. Tosi for very
useful discussions. 
We thank S. Desidera and M. Montalto for information about membership of star
NGC~6253-2508.
This research has made use of the WEBDA database, created by J.-C. Mermilliod
and now operated at the Institute for  Astronomy of the University of Vienna,  
and of data products from the Two Micron All Sky Survey, which is a joint
project of the University of Massachusetts and the Infrared Processing and
Analysis Center/California Institute of Technology, funded by the National
Aeronautics and Space Administration and the National Science Foundation.  
This work was partially funded by the Italian MIUR under PRIN 2003-029437. 

\end{acknowledgements}

\end{document}